\documentclass[useAMS,usegraphicx]{mn2e}
\usepackage{psfig}   
\usepackage{graphicx}
\usepackage{subfigure}

\newcommand{\aap}{A{\&}A}

\newcommand{\mnras}{MNRAS}

\title[What does a universal IMF imply about star formation?]{What does a universal IMF imply about star formation?}
\author[S.P. Goodwin \&
  M.B.N. Kouwenhoven]{Simon~P.~Goodwin\thanks{E-mail:
    s.goodwin@sheffield.ac.uk} and M.B.N.~Kouwenhoven \\ 
  Department of Physics \& Astronomy, University of
  Sheffield, Hicks Building, Hounsfield Road, Sheffield, S3 7RH, UK}

\begin{document}

\date{}

\pagerange{\pageref{firstpage}--\pageref{lastpage}} \pubyear{2008}

\maketitle

\label{firstpage}

\begin{abstract}

  We show that the same initial mass function (IMF) can result from
  very different modes of star formation from very similar underlying
  core and/or system mass functions.  In particular, we show that the
  canonical IMF can be recovered from very similar system mass
  functions, but with very different mass ratio distributions within
  those systems.  This is a consequence of the basically log-normal
  shapes of all of the distributions.  We also show that the
  relationships between the
  shapes of the core, system, and stellar mass functions may not be
  trivial.  Therefore, different star formation in different
  regions could still result in the same IMF.
\end{abstract}


\section{Introduction}

Finding the form of the mass function of stars, in particular the
initial mass function (IMF), has been
a major goal of stellar and galactic astrophysics since Salpeter's
(1955) seminal study (also see Kroupa 2002; Chabrier 2003; Bonnell et
al. 2007).  The origin of the IMF has recently been a subject of 
intense interest 
(see Bonnell et al. 2007 for a review).  Observationally, studies of
the core mass function (CMF) have shown it to have a similar form to
the IMF (Motte et al. 1998; Testi \& Sergent 1998; Johnstone et
al. 2000,2001; Motte et al. 2001; Johnstone \& Bally 2006; Young et
al. 2006; Alves et al. 2007; Nutter \& Ward-Thompson 2007; Simpson et
al. 2008; Enoch et al. 2008) suggesting a link between the two (see,
in particular, Alves et al. 2007; Goodwin et al. 2008).

It is often assumed that if the IMF of two regions is the same, then
star formation in those two regions must have been basically the
same.  However, the IMF is the mass function of {\em individual}
stars.  Therefore the IMF -- taken in isolation -- ignores the fact
that many stars are in multiple systems, and that most/many stars are
thought to have formed as multiples (e.g., Goodwin \& Kroupa 2005;
Kouwenhoven et al. 2005, 2007; Goodwin et al. 2007; Goodwin et
al. 2008, and references therein; see also Lada 2006). The IMF thus 
ignores a large amount
of information related to the star formation process that is stored in
the binary population.

During star formation, many cores must collapse and fragment into a multiple
system (Goodwin \& Kroupa 2005; Goodwin et al. 2007).  Therefore, 
it is the system mass function (SMF) that we
should expect to follow the CMF (as emphasised by Goodwin et
al. 2008).  The IMF is produced by splitting these multiple systems
into their component parts.  How the masses of the individual stars in
the IMF are distributed depends not only upon the SMF, but on the
binary fraction and mass ratio distribution of systems. Moreover,
these quantities may well be mass-dependent.

In this paper we show that very different models for the conversion of
the CMF to the SMF and from the SMF to the IMF can produce very 
similar IMFs.  In Section~2 we outline our method
and the results, and we discuss the implications in Section~3.

\section{From the CMF to the IMF}

\subsection{Method}

Following Goodwin et al. (2008) we have constructed a simple model of
multiple star formation from `cores'\footnote{Such cores may be
classical isolated cores (e.g., Ward-Thompson et al. 2007), or just
dense regions in which stars form (i.e. the star forming clumps see in
simulations such as those of Bate 2009a,b).}.  We assume a universal core
mass function with a log-normal form (although as we describe below,
this is not an important assumption).  This form is based on the
observations of Alves et al. (2007) and Nutter \& Ward-Thompson
(2007).  We take the average mass to be $\sim 1 M_\odot$, and the
width to be $\sigma_{{\rm log}_{10} M} \sim 0.5$.  We then randomly
sample cores from this CMF in a Monte Carlo simulation (see also
earlier studies by Larson (1973), Elmegreen \& Mathieu (1983), and
Zinnecker (1983)).

\begin{figure}
\centerline{\psfig{figure=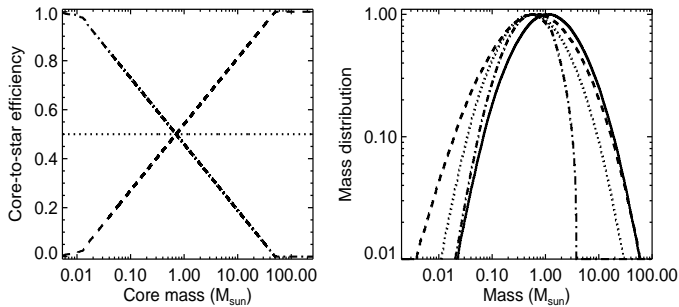,height=4.5cm,width=9.0cm}}
\caption{On the right we show the system mass functions (dotted, dashed, and
  dash-dot lines) produced from a single core mass function (solid
  line), produced by different core-to-star efficiencies with mass
  shown on the left (with corresponding dotted, dashed, and
  dash-dot lines).}
\label{fig:cmf-smf}
\end{figure}

\begin{figure}[H]
\centerline{\psfig{figure=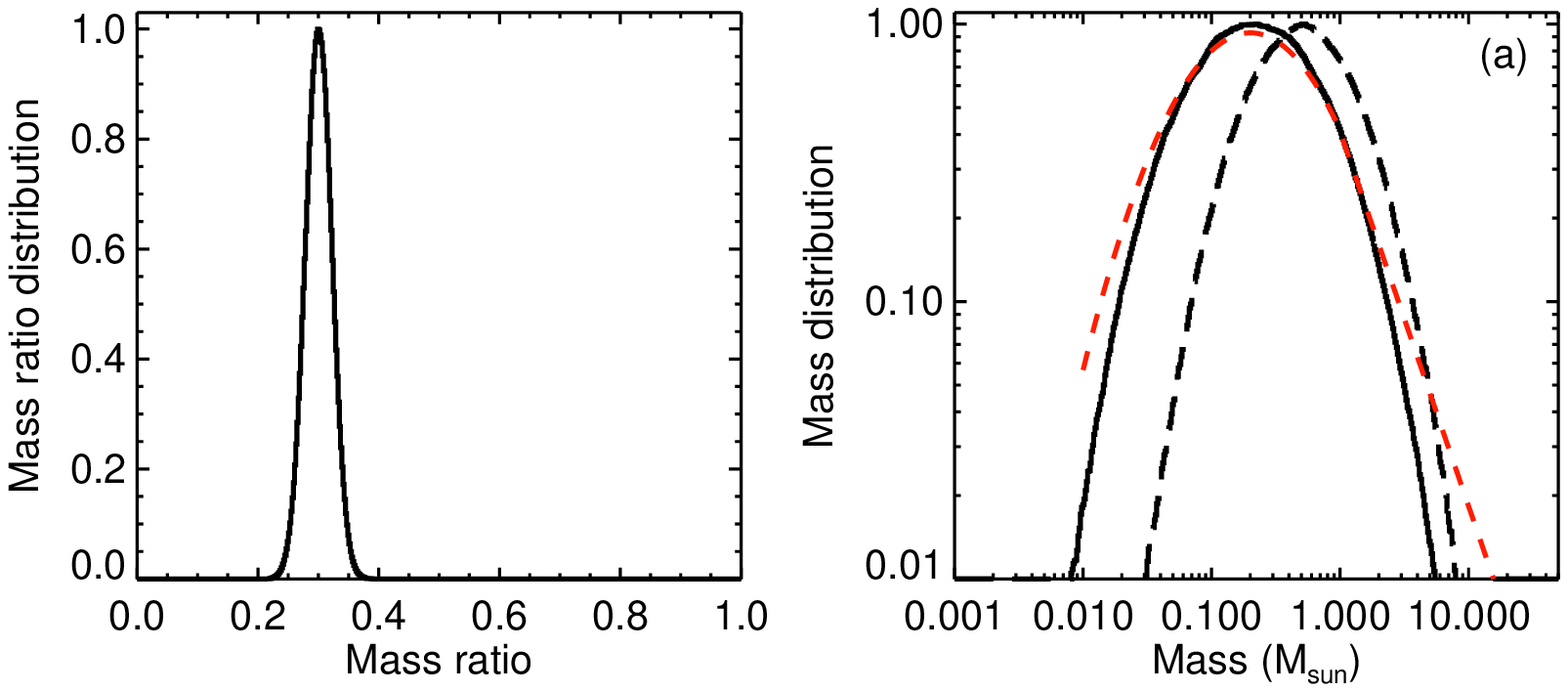,height=4.5cm,width=9.0cm}}
\centerline{\psfig{figure=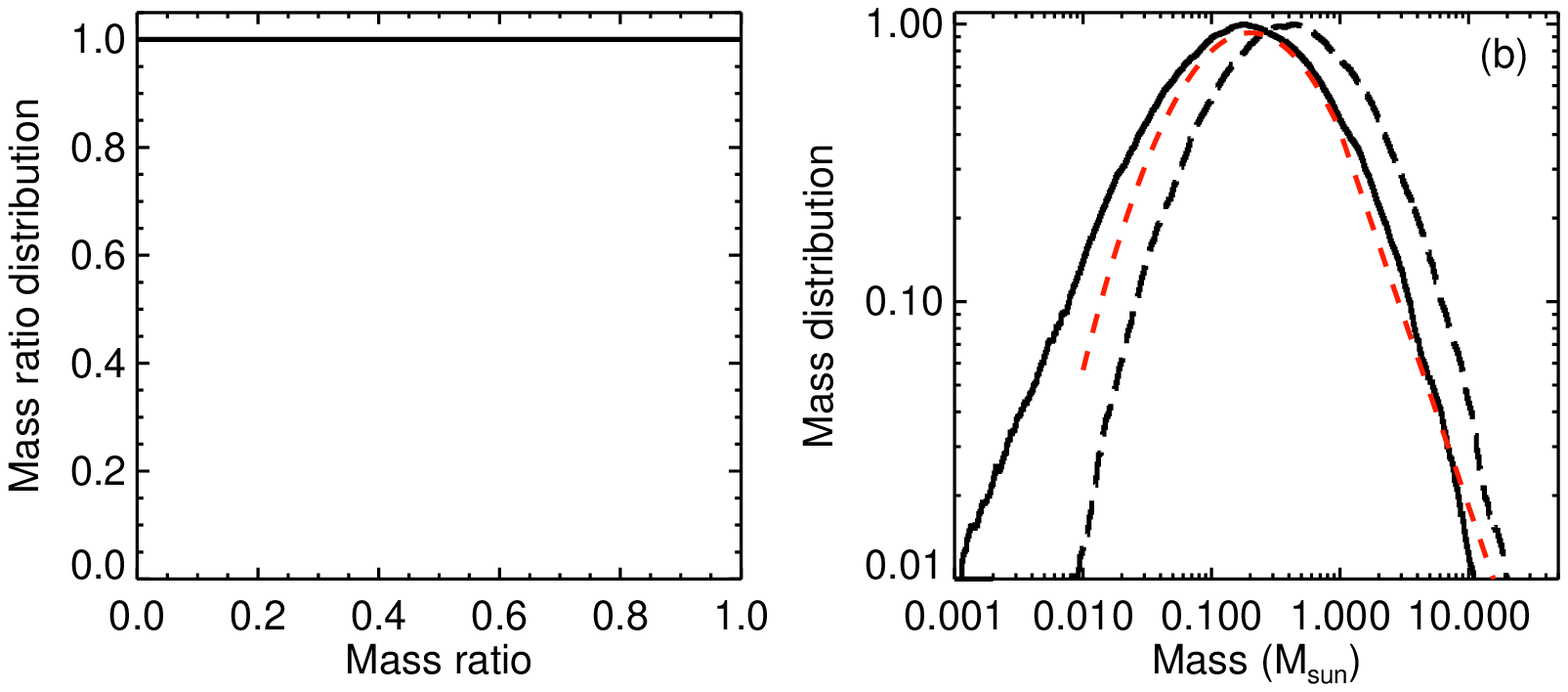,height=4.5cm,width=9.0cm}}
\centerline{\psfig{figure=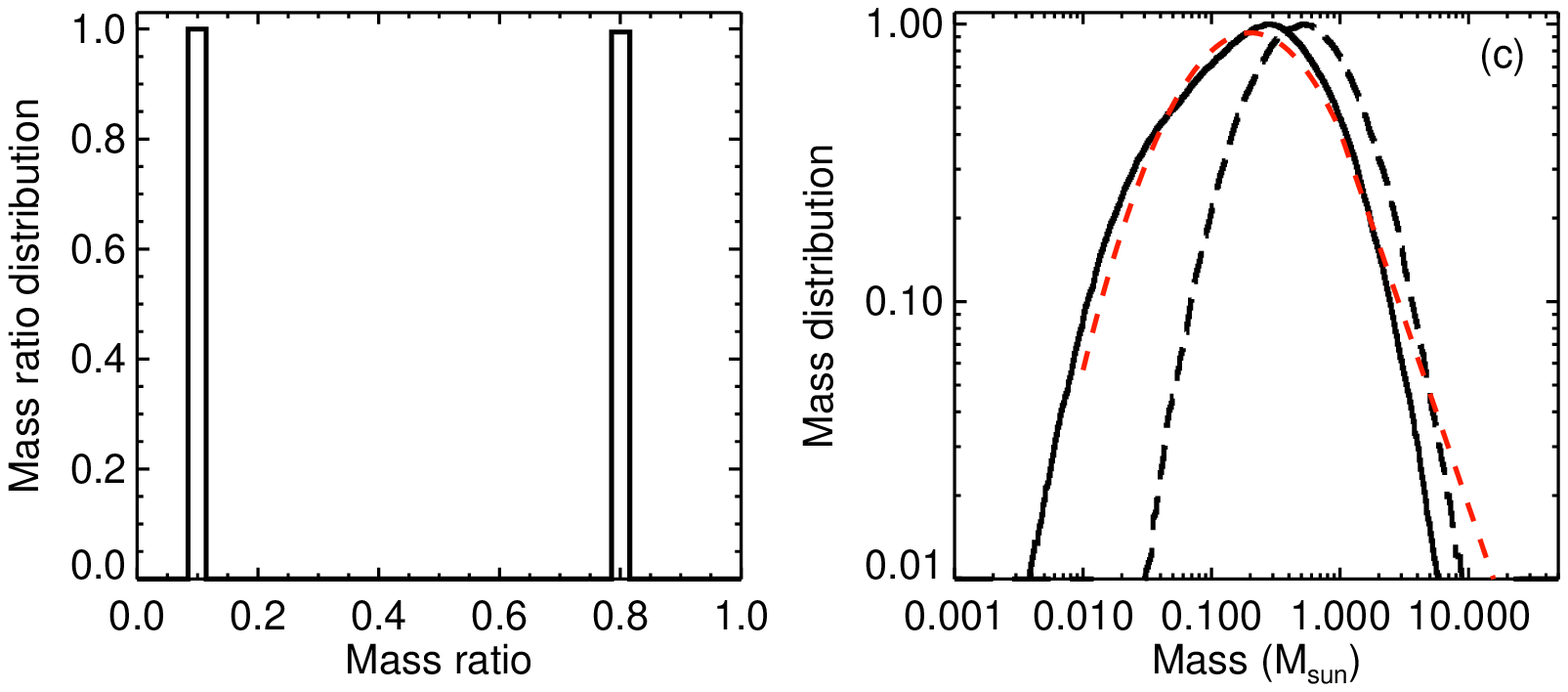,height=4.5cm,width=9.0cm}}
\centerline{\psfig{figure=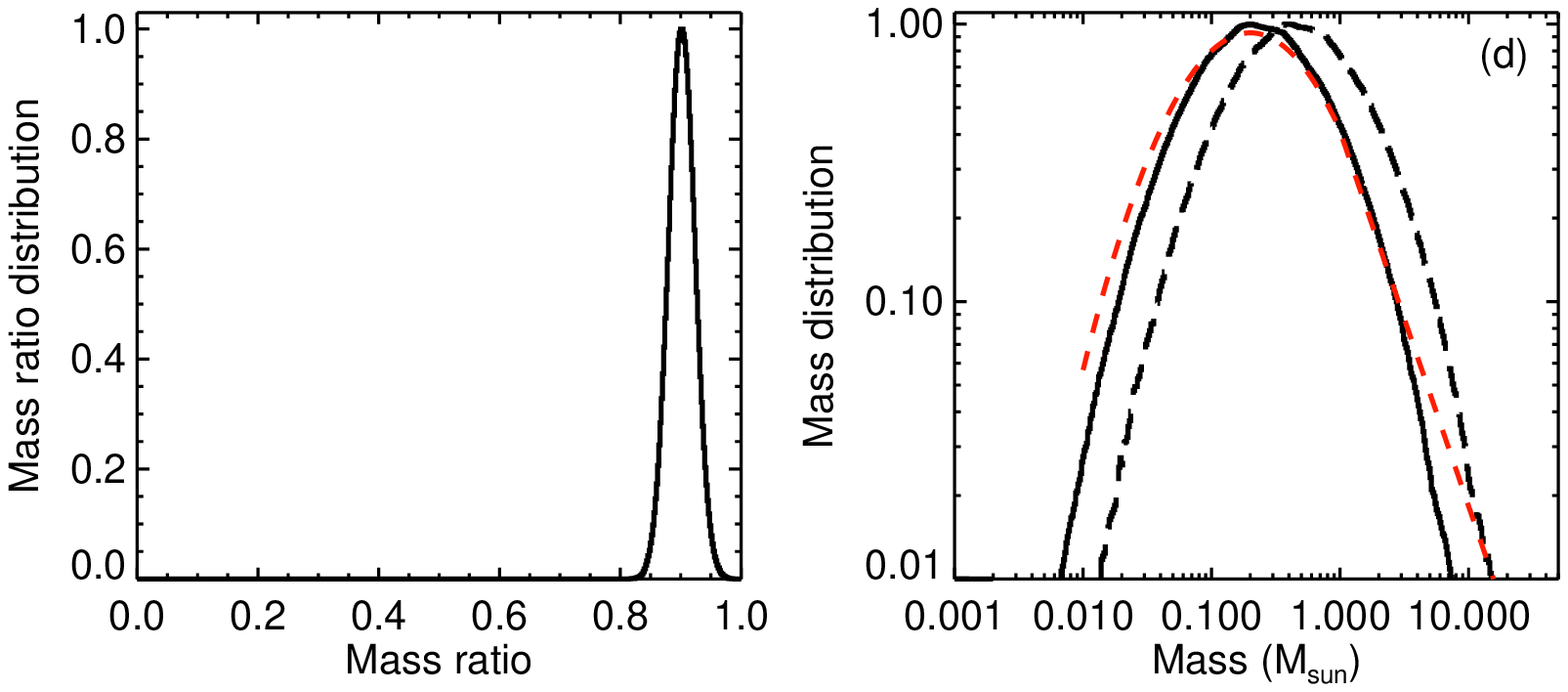,height=4.5cm,width=9.0cm}}
\caption{On the right we show the IMFs (solid black lines) formed from
a given system mass function (dashed line) with a mass ratio
distribution as shown in the left panel assuming a binary fraction of 
unity. The IMFs are compared to a
canonical Chabrier (2003) IMF (light red dashed lines).  The means and
variances of the system mass functions that give the best fits are
(from top to bottom) $\mu_{\rm log_{10} M} = -0.3$, $-0.4$, $-0.3$, and
$-0.35$ and $\sigma_{\rm log_{10} M} = 0.4$, $0.55$, $0.4$,
and $0.5$, respectively.  }
\label{fig:smf-imf}
\end{figure}

We note that taking a log-normal form means that we fail to reproduce
the power-law tail at high-mass that is observed and expected
theoretically (e.g. Padoan \& Nordlund 2002; Hennebelle \& Chabrier
2008), and so we do not expect to fit the IMF properly at high masses.

We create a system mass function from a core mass function by
converting each core into a system with a particular efficiency.
Thus, a core of mass $M_C$ will become a system of mass $M_S =
\epsilon M_C$, where $\epsilon$ is the core-to-star efficiency.  The
core-to-star efficiency may depend on the mass of the core, and even
upon the mass ratio of the stars within the core.

We show below that, depending on the core mass function and the 
core-to-star efficiency, many different system mass functions can be
formed.  Therefore, we use an assumed system mass function and 
then convert the mass of the system $M_S$ into
stars of mass $M_1$ and $M_2$, or or into a single star, depending on
the binary fraction to create an IMF. The mass ratio 
$q=M_2/M_1$ is drawn from a mass
ratio distribution (MRD) in the range $0<q\leq 1$. The corresponding
masses of the primary and companion star are then $M_1 =
M_S(q+1)^{-1}$ and $M_2 = M_S(q^{-1}+1)^{-1}$ (see Kouwenhoven et
al. 2008 for a detailed study of pairing functions in binary systems).
For simplicity, we will only consider single stars and binary systems.

Finally, the IMF is given by the distribution of masses of all {\em
individual} stars (single stars, primaries, companions) in all of the
systems.  We then compare this derived IMF to the `canonical' Chabrier
(2003) log-normal IMF.

In summary, we have an initial core mass function (CMF).  A combination 
of the CMF and the core-to-star efficiency (CSE) gives the 
system mass function (SMF).  Combining the SMF and the mass ratio 
distribution (MRD) then gives the stellar initial mass function
(IMF).  Note that the CSE may depend on the MRD as we shall discuss
below.  

We have assumed at the first step that the CMF is universal (and has a
log-normal form).  Clearly this assumption may well be wrong.
Firstly, determinations of the CMF are very difficult, and it is not
clear if the observed CMF is indeed the CMF that should be used as the
underlying distribution from which stars form (Clark et al. 2007;
Hatchell \& Fuller 2008; Smith et al. 2008; Swift \& 
Williams 2008).  However, for the
purposes of this paper we wish to show that {\em even if} the CMF is
universal, we are able to produce the canonical IMF through 
different modes of star formation.  However, as we shall see later, the
  assumed initial log-normal form of the CMF is one of the main reasons why
  the following distributions also maintain a log-normal form.

\subsection{From the CMF to the SMF}

In Fig.~1 we show how it is possible to change the width and shape of 
the SMF formed from the CMF by changing how the CSE varies 
with core mass.  If high-mass cores are more
efficient at converting gas to stars than low-mass cores, the the SMF
is broader than the CMF.  However, if low-mass cores are more
efficient, then the SMF is narrower than the CMF.  But note that the
resultant SMF always has a roughly log-normal form, often with a 
just a small degree of skewness added.

Presumably, the CSE is not independent of mass. The higher the level
of feedback from stars the lower we might expect the CSE to be.
However, this may not be a trivial relationship.  More massive stars
presumably produce more feedback\footnote{Assuming a main
sequence-like relationship the mass-luminosity relationship would go
as $L \propto M^{3.5}$, however the mass-luminosity relationship for
PMS stars is highly complex and age- and mass-dependent.}, and so be
able to reduce the fraction of a core that accretes onto the
system. On the other hand, a more massive core is also more bound and
so more difficult to disperse.

An interesting possibility is that the CSE depends not just on the
mass of the system, but also on the mass ratio of that system.  It is
possible that an unequal-mass system (say, 1.8$M_\odot$--0.2$M_\odot$) 
will produce far more feedback than an equal-mass system (say, 
1$M_\odot$--1$M_\odot$) with the same system mass.  Therefore, 
a larger core is required to
form the first system.  Also, as pointed-out by Myers (2008), the CSE
should also depend on the core density.

\begin{figure}
\centerline{\psfig{figure=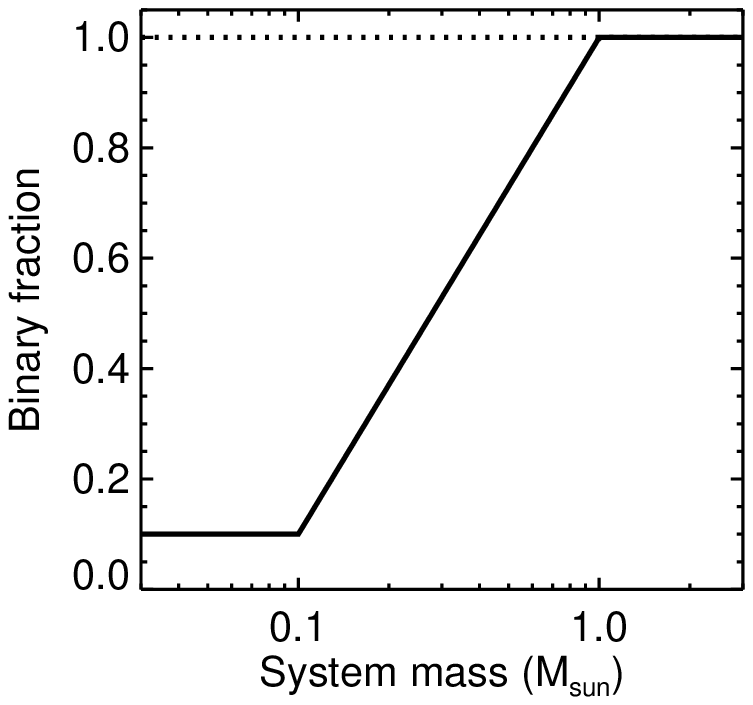,height=4.5cm,width=6.0cm}}
\caption{The dependence of binary fraction on system mass used to
  generate the IMFs from the SMFs in Fig.~2 (dashed-line) -- where all
  systems are binaries -- and, Fig.~4 (solid-line) -- where there is 
  an increasing binary fraction with system mass.}
\label{fig:binfrac}
\end{figure}

\begin{figure}
\centerline{\psfig{figure=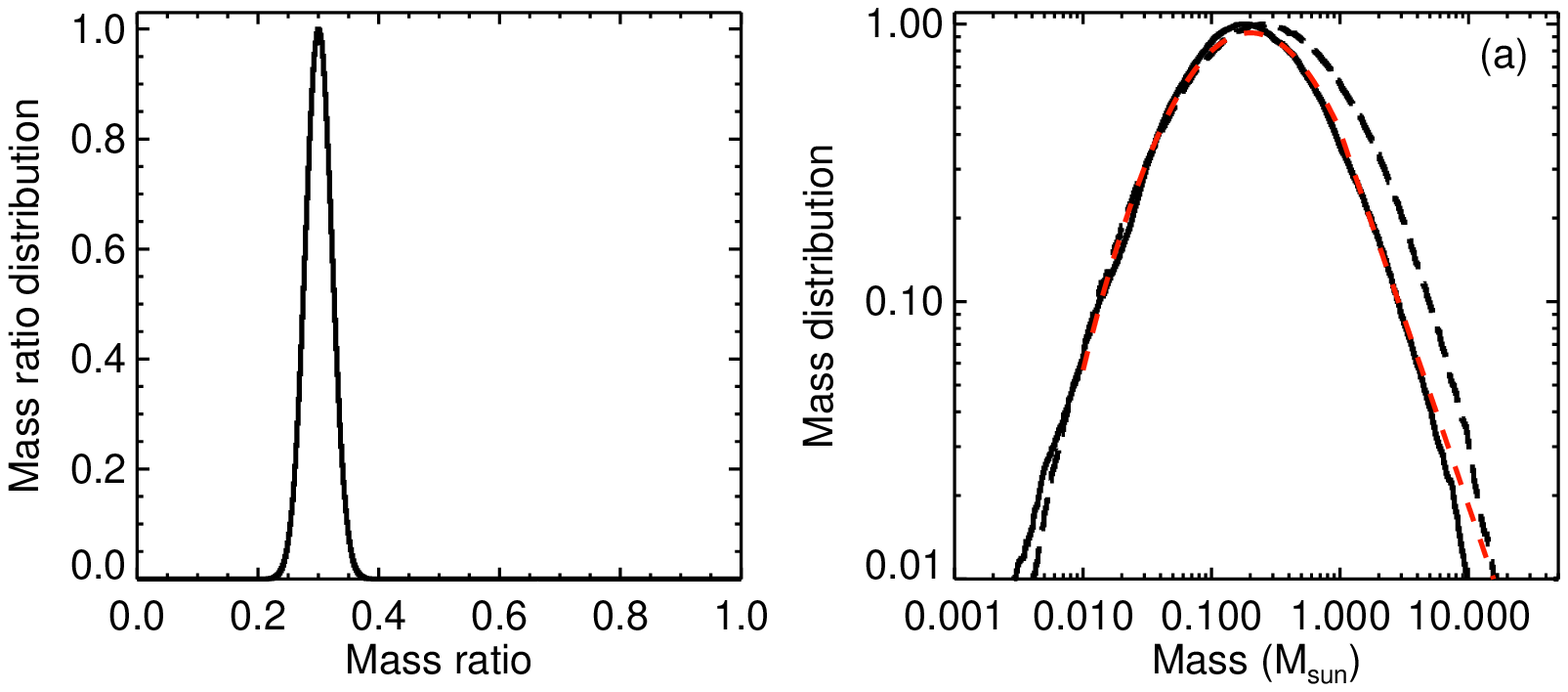,height=4.5cm,width=9.0cm}}
\centerline{\psfig{figure=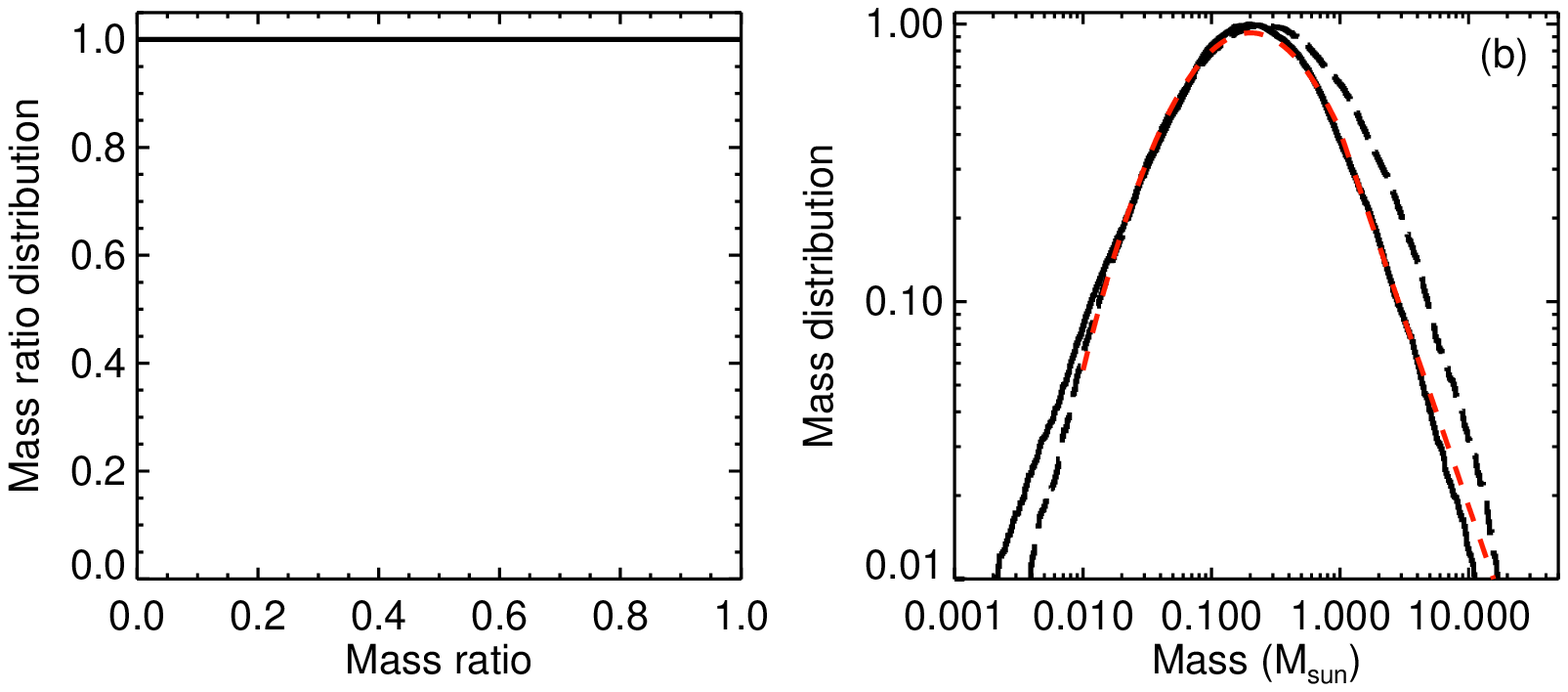,height=4.5cm,width=9.0cm}}
\centerline{\psfig{figure=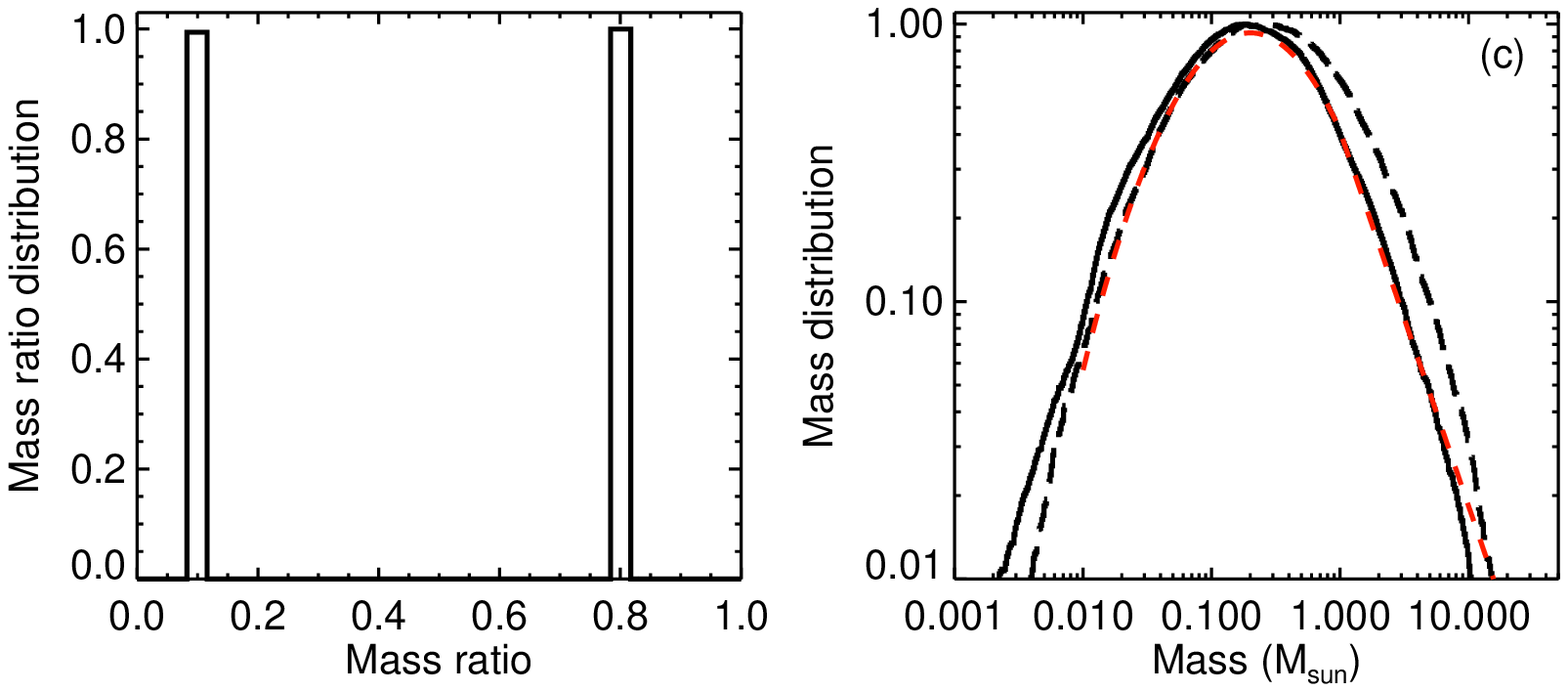,height=4.5cm,width=9.0cm}}
\centerline{\psfig{figure=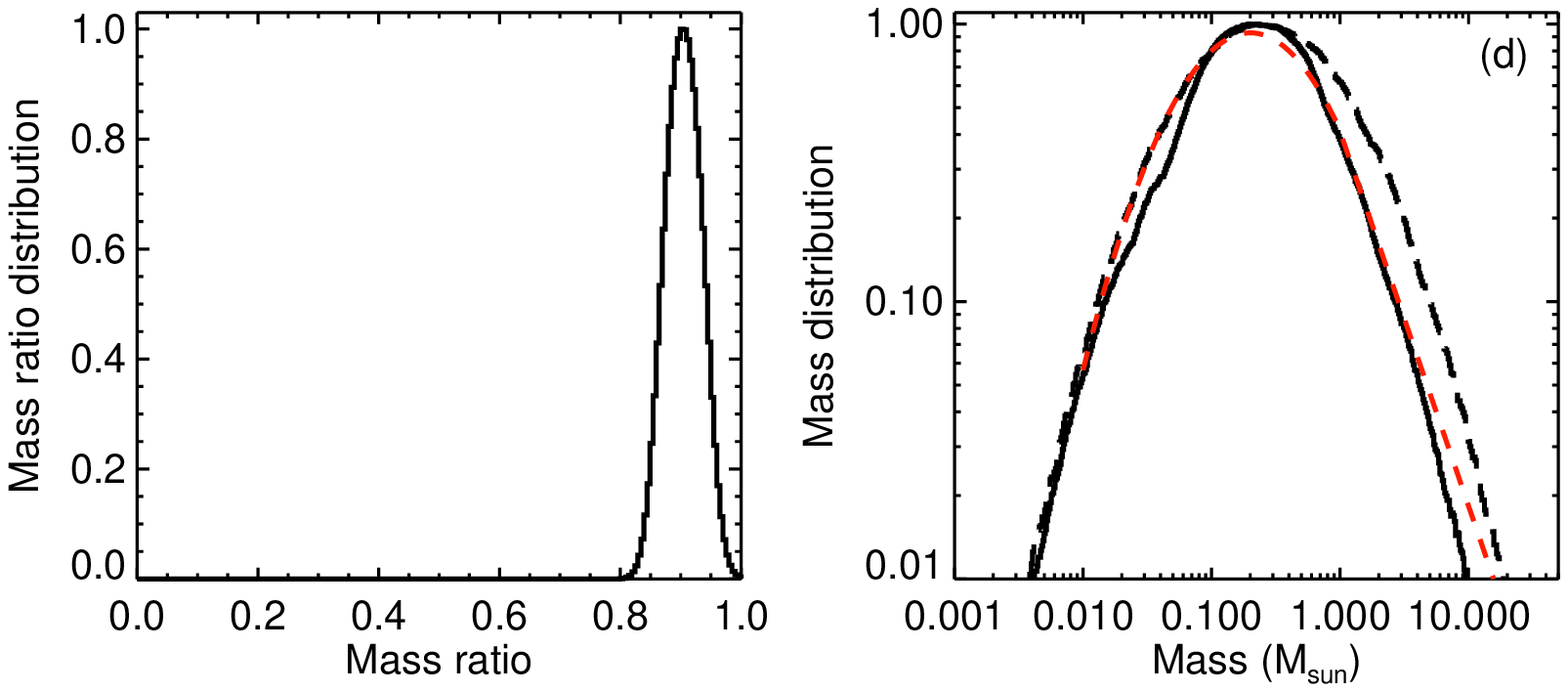,height=4.5cm,width=9.0cm}}
\caption{As Fig.~2 but for a binary fraction that increases with
  system mass 
as shown in Fig.~3 (solid line).  In all cases, the canonical 
IMF can be fitted with an SMF of mean $\mu_{\rm log_{10} M} = -0.6$
and variance $\sigma_{\rm log_{10} M} = -0.6$.}
\label{fig:smf-bin}
\end{figure}

\subsection{From systems to stars}

In Fig.~2 we present the IMFs that result from systems with a binary
fraction of unity for a selection of very different choices for the 
MRD.  We have fine-tuned the SMFs in order to obtain good fits to the
canonical IMF.  What is clear from Fig.~2 is that very different 
choices for the MRD are {\em all} able to produce the canonical IMF.

A potential problem is that we have introduced a fine-tuning element 
in that the SMF has been chosen to give the best-fit to the canonical
IMF for a given MRD.  However, this fine-tuning is not significant.
As indicated in the captions to each part of Fig.~2, the SMFs
do not vary very much, with means of $\mu_{\rm log_{10} M} = -0.3$,
$-0.4$, $-0.3$, and $-0.35$ and variances of $\sigma_{\rm
  log_{10} M} = 0.4$, $0.55$, $0.4$, and $0.5$, respectively, for
the different models.  

It would be extremely difficult, if not impossible, to distinguish
observationally differences in the mean of the SMF between $0.4$ 
and $0.5 M_\odot$, and differences in the
$1 \sigma$ widths of $0.1$ -- $0.2 M_\odot$ at one end, and $1.25$ --
$1.40 M_\odot$ at the other.  Indeed, the models presented in
Figs.~2a and~2c have identical SMFs despite having completely
different MRDs (a single peak at $q \sim 0.3$, compared to peaks at $q
=0.1$ and $0.8$) and yet both give good fits to the canonical IMF.

All of the models presented in Fig.~2 have assumed that {\em all}
systems produce a binary (as shown in Fig.~3).  Clearly, the 
canonical IMF can also be recovered
from a SMF of entirely single stars if the SMF is exactly the same as
the IMF.  However, we know that a significant fraction of systems in
the field are multiples, and that presumably many more were multiples
at birth as multiples are destroyed, but not created (Kroupa 1995a,b; 
Parker et al. 2009).

Goodwin et al. (2008) found that -- assuming a constant CSE and a uniform
MRD -- a model with a binary fraction of unity was a better fit to the
canonical IMF than one in which binarity declined with system
(ie. primary) mass as is observed in the field (see Lada 2006).

In Fig.~4 we show the best fits to the canonical IMF for the same set
of MRDs as in Fig.~2, but for a binary fraction that varies with
system mass as shown in Fig.~3 -- unity for a system mass $>1
M_\odot$, linear for $0.1 < M_{\rm sys}/M_\odot < 1$, and $0.1$ for
$M_{\rm sys} < 0.1 M_\odot$.  In order to fit the canonical IMF for each
MRD the best SMFs are $\mu_{\rm log_{10} M} = -0.6$ and
$\sigma_{\rm log_{10} M} = 0.6$ for {\em all} of our models.  The 
peaks of the SMFs ($\sim 0.25
M_\odot$) are lower than for a binary fraction of unity as the reduced
number of low-mass companions means that the low-mass end of the IMF
must be made-up largely of single stars directly from the SMF.  The
increased width is then required in order to allow sufficient numbers
of higher-mass stars to form.

It might seem that this result is at odds with Goodwin et al. (2008).
However, Goodwin et al. set the CMF with which to generate the SMF and
then the IMF as being the Orion/Pipe Nebula CMF (with a mean of $\sim
1 M_\odot$ and variance of $\sigma_{\rm log_{10} M} = 0.55$, see 
above) and a constant CSE.  Given these constraints it is impossible 
to form an SMF with a mean
of $0.25 M_\odot$ and a variance of $\sigma_{\rm log_{10} M} = 0.6$
as required to
produce the canonical IMF with a decreasing binary fraction.  However,
as we show here, if the CMF is different, or if the CSE is such that
the variance is increased and the mean lowered significantly such a 
CMF can produce the canonical IMF with a varying binary fraction.

We have also examined the effect of using a CMF that is not
  log-normal (in particular, triangular and top-hat CMFs).  Features
  in the CMF are always reflected in some way (even if distorted) in
  the SMF and the IMF no matter our choice of the CSE and MRD.  Thus
  the apparently smooth
  log-normal-like shape of the IMF would seem to be a natural
  consequence of the log-normal-like shape of the CMF (cf. Elmegreen
  \& Mathieu 1983; Zinnecker 1983).  However, the lack of features 
  tells us that the CMF that produces the SMF that produces the
  IMF is highly unlikely to contain any sharp features or
  discontinuities as these should be apparent in the IMF.

It is surprising that the IMF is so insensitive to the MRD of
  stars.  The reason is that the IMF is so insensitive is that it
  retains the log-normal-like shape of the SMF (which itself is
  retained from the CMF).  As the SMF is distributed in a
  roughly log-normal way, each chosen mass ratio from the MRD retains
  this shape  (e.g. for an MRD of two delta functions, the SMF will
  produce two stellar distributions, each a log-normal).  Thus the IMF
  is the sum of a number of log-normals thus retaining the log-normal
  shape\footnote{It is possible to produce other distributions.  For
  example an MRD which always produces a star with 99.9 per cent of the
  mass in a core, and a planet with 0.1 per cent of the mass.  In such
  a situation the two log-normals would be so far apart that they
  would appear as two separate log-normals.  However observations show
  that typical mass ratios are in the range $0.1$ -- $1$
  (e.g. Duquennoy \& Mayor 1991).}.

It is interesting to examine what form we might expect the MRD to
  have.  We would expect that most binaries form via the fragmentation
  of discs (or massive disc-like objects) around young stars (see
  Goodwin et al. 2007 for a review).  If discs fragment early then the
  secondary will form whilst there is still a significant amount of
  material to accrete and we might expect the MRD to favour more 
  equal-mass systems.  If the discs fragment late after the primary
  has accreted most of its mass then there will be little material
  left to accrete and unequal-mass systems might predominate (as might 
  also happen if the secondary formed at a large distance from the
  primary).  Thus the MRD might contain information on the formation
  time of secondaries (which might depend on the amount of angular
  momentum or strength of magnetic fields in the core?).  If the MRD
  favours unequal-mass binaries then the primary IMF will reflect well
  the form of the SMF, however if many binaries are more equal-mass
  then the primary IMF would not.

\section{Discussion and Conclusions}

We have investigated the origin of the stellar initial mass function
(IMF) as produced by a system mass function (SMF), which is itself
formed from a particular core mass function (CMF).  The CMF is
converted to the SMF via a core-to-star efficiency (CSE) factor, and
then turned into the IMF via a mass ratio distribution (MRD).  Our
main findings are:
\begin{itemize}
\item Different core-to-star efficiencies can significantly change
  the width and shape of the system mass function formed from a given
  core mass function.
\item Very different mass ratio distributions within systems can
  produce very similar (and canonical) IMFs
  from very similar system mass functions.
\end{itemize}
Thus, we conclude that observing the same initial mass function in
different regions does not necessarily mean that star formation 
was the same in those regions.  For very different choices of the mass 
ratio distribution in star forming cores, the same (canonical) IMF may
be found.  In addition, the form of the IMF is surprisingly
insensitive to the variation of the binary fraction with mass (modulo
the position of the peak of the system mass function).

This raises a question about the universality, or non-universality, of
  star formation based on observing the same IMF in different regions.
If one star forming region produced systems with generally equal-mass
stars, whilst another produced systems with generally comparatively
low-mass companions, it seems very difficult to claim that star
formation in those regions was the same.  However, the core mass
functions, system mass functions, and the IMFs could all be very
similar.  Therefore, in order to claim that star formation is the same
in different regions, the {\em birth} system mass function and mass
ratio distribution of binaries, and the binary fraction, must also be
the same.

It is important to note that it is the {\em birth} system mass
functions and mass ratio distributions that must be determined.
Dynamical evolution can seriously alter the initial binary population
in a cluster (e.g. Kroupa 1995a, 1995b; Parker et al. 2009) and destroy much
of the vital information on the birth properties.  

It is important to note that the peak mass of the IMF which
  always seems to occur at $\sim 0.1$ -- $0.3 M_\odot$ does constrain
  the star formation process.  The initial core mass functions must always have a
  similar peak mass which is then imposed on the system mass function
  (regulated by the core-to-star efficiency) which is then imposed on
  the IMF (but regulated by the mass ratio distribution including the
  binary fraction).  If the binary fraction varies between different
  star formating regions the peak of the core mass functions could
  change by a factor of $2$ or $3$, but no more.  Thus the underlying
  core mass functions in all regions should probably have a peak mass
  of order $1 M_\odot$.

\section{Acknowledgements}
We would like to thank the referee, Charlie Lada, for his useful
comments.  M.B.N.K. was supported by PPARC/STFC under grant 
number PP/D002036/1. We
acknowledge networking funding from the British Council under the
Prime Minister's Initiative 2 through grant No. RC01.

\end{document}